\documentclass[letterpaper,twocolumn,prl,showpacs]{revtex4-1}

\usepackage{bm}
\usepackage{amsmath}
\usepackage{amssymb}
\usepackage[hypertex]{hyperref}
\usepackage{graphics}
\usepackage{graphicx}

\allowdisplaybreaks[1]

\newcommand{\beq}{\begin{equation}}
\newcommand{\eeq}{\end{equation}}
\newcommand{\bea}{\begin{eqnarray}}
\newcommand{\eea}{\end{eqnarray}}
\newcommand{\beano}{\begin{eqnarray*}}
\newcommand{\eeano}{\end{eqnarray*}}
\newcommand{\no}{\nonumber}

\newcommand{\ra}{\rangle}
\newcommand{\la}{\langle}

\newcommand{\al}{\alpha}
\newcommand{\be}{\beta}

\newcommand{\up}{\uparrow}
\newcommand{\down}{\downarrow}
\newcommand{\dn}{\downarrow}
\newcommand{\bup}{\bph_\up}
\newcommand{\fup}{\fph_\up}
\newcommand{\bdup}{\bd_\up}
\newcommand{\fdup}{\fd_\up}
\newcommand{\bdown}{\bph_\down}
\newcommand{\fdown}{\fph_\down}
\newcommand{\bddown}{\bd_\down}
\newcommand{\fddown}{\fd_\down}
\newcommand{\bbup}{\bbph_\up}
\newcommand{\fbup}{\fbph_\up}
\newcommand{\bbdup}{\bbd_\up}
\newcommand{\fbdup}{\fbd_\up}
\newcommand{\bbdown}{\bbph_\down}
\newcommand{\fbdown}{\fbph_\down}
\newcommand{\bbddown}{\bbd_\down}
\newcommand{\fbddown}{\fbd_\down}

\newcommand{\bb}{\bar{b}}
\newcommand{\fb}{\bar{f}}

\newcommand{\cb}{\bar{c}}

\newcommand{\bd}{b^{\dagger}}
\newcommand{\fd}{f^{\dagger}}

\newcommand{\cd}{c^{\dagger}}

\renewcommand{\cup}{\cph_\up}
\newcommand{\cdup}{\cd_\up}
\newcommand{\cdown}{\cph_\down}
\newcommand{\cddown}{\cd_\down}

\newcommand{\cph}{c^{\vphantom{\star}}}

\newcommand{\bph}{b^{\vphantom{\dagger}}}
\newcommand{\fph}{f^{\vphantom{\dagger}}}

\newcommand{\fbph}{{\bar f}^{\vphantom{\dagger}}}
\newcommand{\bbph}{{\bar b}^{\vphantom{\dagger}}}

\newcommand{\bbd}{\bb^{\dagger}}
\newcommand{\fbd}{\fb^{\dagger}}

\newcommand{\cbd}{\cb^{\dagger}}

\newcommand{\Jb}{\bar J}
\newcommand{\Kb}{\bar K}
\newcommand{\Bb}{\bar B}
\newcommand{\Qb}{\bar Q}
\newcommand{\Vb}{\bar V}
\newcommand{\Wb}{\bar W}

\newcommand{\Aph}{A^{\vphantom\dagger}}

\begin{document}
\bibliographystyle{plainnat}

\title{Relevant perturbations at the spin quantum Hall transition}

\author{Shanthanu Bhardwaj}
\affiliation{Department of Physics, The University of Chicago, Chicago, Illinois 60637, USA}

\author{Ilya A. Gruzberg}
\affiliation{Department of Physics, The Ohio State University, Columbus, OH 43210, USA}

\author{Victor Kagalovsky}
\affiliation{Shamoon College of Engineering, Beer-Sheva 84105, Israel}

\date{January 27, 2014}

\begin{abstract}

We study relevant perturbations at the spin quantum Hall critical point using a network model formulation. The model has been previously mapped to classical percolation on a square lattice, and we use the mapping to extract exact analytical values of the scaling dimensions of the relevant perturbations. We find that several perturbations that are distinct in the network model formulation correspond to the same operator in the percolation picture. We confirm our analytical results by comparing them with numerical simulations of the network model.

\end{abstract}

\pacs{72.15.Rn, 73.20.Fz, 73.43.-f}

\maketitle

\section{Introduction}
\label{sec:introduction}

Anderson localization of a quantum particle \cite{Anderson58} or a classical wave in a random environment is a vibrant research field \cite{AL50}. One of its central research directions is the physics of Anderson transitions \cite{evers08}, quantum critical points tuned by disorder. These include metal-insulator transitions and transitions of quantum Hall type separating distinct phases of topological insulators. While such transitions are conventionally observed in electronic (metallic and semiconductor) structures, there is also a considerable number of other experimental realizations actively studied in recent and current works. These include localization of light \cite{wiersma97} and microwaves \cite{microwaves}, cold atoms \cite{BEC-localization} (see a recent review [\onlinecite{Shapiro-review-2012}]), ultrasound \cite{faez09}, and optically driven atomic systems \cite{lemarie10}. 

From the theoretical point of view, symmetries play a central role in determination of universality classes of critical phenomena. This idea was applied to Anderson localization by Altland and Zirnbaueer (AZ) \cite{AltlandZirnbauer} who identified ten distinct symmetry classes. In three of these classes, classes A, C, and D in AZ classification, the time-reversal invariance is broken, and there is a possibility for a quantum Hall transition in two dimensions.

The transition in class A is the usual integer quantum Hall (IQH) transition in a two-dimensional (2D) electronic system in a strong perpendicular magnetic field (see Ref. [\onlinecite{Huckestein:95}] for a review). Class A also includes the model of electrons in a random magnetic field, where all states are believed to be localized \cite{LeeChalker}.

Class C is one of the four Bogolyubov-de Gennes classes which describe transport of quasiparticles in disordered superconductors at a mean field level, and possess the particle-hole symmetry. In this class the spin-rotation invariance is preserved, the quasiparticles have conserved spin, and one can study spin transport. The corresponding Hall transition is known as the spin quantum Hall (SQH) transition \cite{Kagalovsky:99, SenthilMarstonFisher:99}, at which the system exhibits a jump in the spin Hall conductance from 0 to 2 in appropriate units.

In spite of tremendous efforts, most models of Anderson transitions have resisted analytical treatment. The IQH transition is one prominent example where only recently some analytical progress has been achieved \cite{BettelheimGruzbergLudwig}. On the other hand, the SQH transition enjoys a special status, since a network model of this transition was mapped exactly to classical percolation on a square lattice \cite{GruzReadLudwig:99}. The original mapping used the supersymmetry (SUSY) method of Efetov \cite{Efetov} adapted to networks \cite{Read, GRS:97}. An alternative way to obtain the mapping was found later \cite{BeamondCardyChalker02, EversMildenbergerMirlin}. It was also extended to network models in class C on arbitrary graphs \cite{Cardy:2005}. Many exact results are known for classical percolation. Thus, the mapping has lead to a host of exact critical properties at the SQH transition \cite{GruzReadLudwig:99, EversMildenbergerMirlin, Cardy:2000, Cardy:2005, Subramaniam-et-al:06}. However, these results are not exhaustive, since not all possible relevant perturbations were considered in Refs. [\onlinecite{GruzReadLudwig:99, EversMildenbergerMirlin,  Cardy:2000, Subramaniam-et-al:06}]. Several critical exponents have been obtained numerically in Refs. [\onlinecite{Kagalovsky:99, SenthilMarstonFisher:99, Kagalovsky:2001, Kagalovsky:Annalen, Obuse-et-al:10}].

In this paper we reexamine the relevant perturbations at the SQH critical point. As our main tool we use the SUSY method applied to the simplest network model in class C describing the SQH effect. We introduce all possible perturbations that are relevant at the critical point of the SHQ network model. One of them preserves the symmetries of the model and drives the SQH transition. Other relevant perturbations break symmetries specific to class C and lead to a crossover to class A. We use the percolation mapping of Ref. [\onlinecite{GruzReadLudwig:99}] to extract analytical values of the scaling dimensions of all relevant perturbations. As a result, we find that one of the results of Ref. [\onlinecite{GruzReadLudwig:99}] does not hold, and find the correct value of the corresponding critical exponent. In addition, we find that several microscopically distinct perturbations all have the same scaling dimension related to a single operator in the percolation picture.

The paper is organized as follows. In Sec. \ref{sec:model} we describe the network model appropriate for the study of the spin quantum Hall transition in class C and relevant perturbations near it, and summarize our results. In Sec. \ref{sec:SUSY} we briefly describe the SUSY method for the network, and derive the second-quantized supersymmetric transfer matrices. These matrices are then averaged over quenched disorder. In Sec. \ref{sec:superspin-chain} we take an anisotropic limit, thereby mapping the network model to a superspin chain. The superspin chain contains a critical point, and several relevant perturbations. All terms in the superspin chain Hamiltonian are interpreted in terms of the classical percolation picture of Ref. [\onlinecite{GruzReadLudwig:99}], and this interpretation allows us to extract dimensions of all relevant perturbations and the corresponding critical exponents. In Sec. \ref{sec:numerics} we present our recent numerical results, discuss results of other numerical simulations of the network model, and compare all these with our analytical predictions. We then conclude. For completeness, we review details of the SUSY method for the class C network model in a series of Appendices.

\section{The model and a summary of results}
\label{sec:model}

A scattering theory description of Anderson localization and Anderson transitions in terms of random network models was introduced in Ref. [\onlinecite{Shapiro82}]. For systems exhibiting quantum Hall effects one can use semiclassical drifting orbits \cite{Trugman83, Shapiro86} scattered at saddle points of a smooth random potential to provide an intuitive derivation for network models. The resulting networks are chiral, reflecting the breaking of time reversal invariance in strong magnetic fields. The simplest such model is the the Chalker-Coddington (CC) model originally proposed to describe the IQH effect \cite{ChalkerCoddington:88}.

\begin{figure}[t]
\centering \includegraphics[width=0.9\columnwidth]{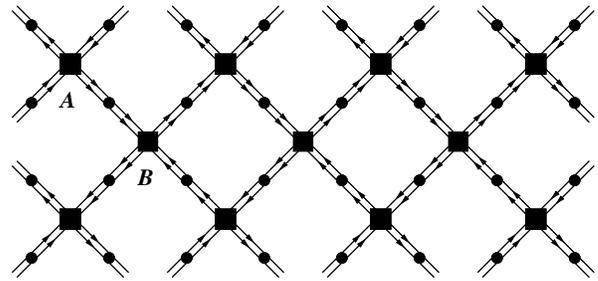}
\caption{Two-channel chiral network model. Dots represent scattering matrices on the links (\ref{link-matrix-U2}) and squares represent the nodal scattering matrices (\ref{node-matrices}).}
\label{Fig:Network}
\vskip -3mm
\end{figure}

Here we consider a generalization of the CC model shown in Fig. \ref{Fig:Network}. In this network each link supports two co-propagating channels which we label by $\sigma = \, \uparrow, \downarrow$. The corresponding doublets of complex fluxes propagate along links, and their components get mixed by scattering matrices ${\cal S}_{\rm link}$, which relate the incoming and outgoing fluxes. In class A, the symmetry class of the IQH effect, the scattering matrices on the links are general unitary U$(2)$ matrices, and can be parameterized as
\begin{align}
{\cal S}_\delta = e^{i\delta} {\cal S}_0,
\label{link-matrix-U2}
\end{align}
where the matrix ${\cal S}_0 \in \text{SU}(2)$. The link matrices are independent identically distributed random variables whose distribution is chosen depending on a specific physical situation.

As in the ordinary CC network there are two sublattices, $A$ and $B$, on which the nodes are related by a 90$^\circ$ rotation.  Scattering of the fluxes at the nodes (black squares) is described by orthogonal matrices diagonal in spin indices: ${\cal S}_S = {\cal S}_{S\up} \oplus {\cal S}_{S\down}$,
\begin{align}
{\cal S}_{S\sigma} &= \begin{pmatrix} r_{S\sigma} & t_{S\sigma}^{\vphantom{2}} \\
-t_{S\sigma}^{\vphantom{2}} & r_{S\sigma} \end{pmatrix}, & r_{S\sigma} &\equiv (1-t_{S\sigma}^2)^{1/2},
\label{node-matrices}
\end{align}
where $S=A$, $B$ labels the sublattice. Usually the scattering amplitudes on the two sublattices $t_{S\sigma}$ are assumed to be non-random. The network has a critical point at
\begin{align}
t_{A\up} = t_{B\up}= t_{A\down} = t_{B\down}.
\label{critical-point}
\end{align}

Depending on the choice of parameters and their probability distributions, the generalized two-channel network model in class A can be used to describe various physical systems: spin-degenerate Landau levels and localization in a random magnetic field \cite{LeeChalker}, the IQH effect in a double-layer system \cite{SorMac,GramRaikh}, and the splitting of delocalized states due to the valley mixing in graphene \cite{OGM}. Using the SUSY method, in Ref. [\onlinecite{BhardwajMkhitaryanGruzberg:14}] we have provided a comparative study of the relevant networks and related models.

Let us now consider class C, the symmetry class of the SQH effect. Similar to previous works, we study the SQH transition in (the mean field description of) a singlet superconductor after a particle-hole transformation on the down-spin particles \cite{Senthil-et-al:98}. The transformation interchanges the roles of particle number and $z$ component of spin, and so particle number is conserved rather than spin. This somewhat obscures the spin-rotation symmetry, but makes it possible to use a single particle description and, in particular, a network model. The single-particle energy $(E)$ spectrum has a particle-hole symmetry \cite{AltlandZirnbauer}, so, when states are filled up to $E = 0$, the positive-energy particle and hole excitations become doublets of the global SU(2) symmetry. In this picture, a uniform Zeeman magnetic field $B_z$ for the quasiparticles maps onto a simple shift in the Fermi energy to $E \propto B_z$ \cite{SenthilMarstonFisher:99}, splitting the degeneracy.

In the network model description a particle of either spin and with $E = 0$ is represented by a doublet of complex fluxes that can propagate in one direction along each link (Fig. \ref{Fig:Network}). The global spin-rotation symmetry of class C requires the scattering matrices to be unitary symplectic. Thus in the two-channel model the link matrices belong to $\text{Sp}(2) \cong \text{SU}(2)$, and in the parametrization (\ref{link-matrix-U2}) we have to set the overall phase $\delta = 0$. The absence of an additional (random or deterministic) U(1) phase here is crucial. Taking the link matrices ${\cal S}_0$ to be uniformly distributed over the Haar measure on SU(2) we obtain the model that maps to a classical bond percolation on the square lattice \cite{GruzReadLudwig:99}. Both the absence of the overall phase {\it and} the uniform distribution over SU(2) are essential technical ingredients of the mapping (as explained in the Appendix).

Let us describe how a non-zero energy $E$ enters the network model description. A state of the network (the collection of the fluxes on all channels joining scattering matrices) evolves in discrete time steps under the action of a unitary evolution operator $\cal U$ which has nonzero matrix elements only between pairs of incoming and outgoing channels scattered on a link or at a node, the matrix elements simply being the scattering amplitudes relating the corresponding fluxes. The main object of study is the Green's function, or the resolvent, of the evolution operator:
\begin{align}
G(e',e;z) = \langle e' |(1 - z{\cal U})^{-1}| e \rangle,
\label{Greens-function}
\end{align}
where $e$ and $e'$ are two channels (edges) of the network. In a closed network $\cal U$ is unitary, and the resolvent has singularities on the unit circle in the complex plain of the spectral parameter $z$. Roughly speaking, if we write ${\cal U} = e^{i{\cal H}}$ and $z = e^{i(E + i \eta)}$, $\cal H$ can be thought of as the Hamiltonian for the network, and $E + i\eta$ as the energy with a finite level broadening $\eta$. The level broadening may be induced by attaching ideal leads that make the network open and break unitarity. Scaling of various observables with energy $E$ close to the critical point is the same as with its imaginary part $\eta$ \cite{EversMildenbergerMirlin, Subramaniam-et-al:06}. This fact allows us to use the real $z = e ^{-\eta}$.

If we expand the Green's function (\ref{Greens-function}) into a power series in $z$, it is clear that a factor of $z$ is associated with each scattering event. In fact, it is sufficient to assign the factors of $z$ only to scattering at the nodes or on the links. We choose the latter option. This leads to the modified link scattering matrices
\begin{align}
{\cal S}_\phi &= e^{i\phi} {\cal S}_0, & \phi &\equiv \delta +i \eta.
\label{link-matrix-eta}
\end{align}
The notation we use stresses the fact that the phase $\delta$ and the level broadening $\eta$ combine to form a single complex parameter $\phi = \delta + i \eta$ where $\delta$ plays the role of the energy $E$. As expected, the modified scattering matrices are not unitary, since a finite level broadening leads to decay of the states of the network and breaks current conservation. We also note here that while $\delta$ can be random, the level broadening $\eta$ will be taken the same for every link.

\begin{table}[t]
  \begin{tabular}{ |l | c | c | c | c | c |}
  \hline
  Exponent & $\nu$ & $\nu_B$ & $\mu_\Delta$ & $\mu_p$ & $\mu_{\delta_0}$ \\ \hline
   \hline
\multicolumn{6}{ |c| }{Analytical predictions} \\
\hline
Ref. [\onlinecite{GruzReadLudwig:99}] & $4/3$ & $4/7$ & $3/2$ & -- & -- \\
         & $\approx 1.33$ & $ \approx 0.57$ & $ = 1.5$ &  &  \\ \hline
This work & -- & -- & $8/7$ & $8/7$ & $4/7$ \\
 &  &  & $ \approx 1.14$ & $ \approx 1.14$ & $ \approx 0.57$ \\ \hline
   \hline
\multicolumn{6}{ |c| }{Numerical results} \\
\hline
   Ref. [\onlinecite{Kagalovsky:99}] & $1.12$ & -- & $1.45$ & -- & -- \\ \hline
   Ref. [\onlinecite{SenthilMarstonFisher:99}] & $1.32(2)$ & $0.55(1)$ & -- & -- & -- \\ \hline
   Ref. [\onlinecite{Kagalovsky:Annalen}] & $1.12$ & -- & $1.45$ & $1.17$ & -- \\ \hline
   Ref. [\onlinecite{Kagalovsky:2001}] & $1.12$ & -- & $1.45$ & -- & $0.7$ \\ \hline
   Ref. [\onlinecite{Obuse-et-al:10}] & $1.33(1)$ & -- & -- & -- & -- \\ \hline
   This work & -- & -- & -- & $1.15$ & -- \\ \hline
   \end{tabular}
   \caption{A summary of previous and new results for critical exponents at the SQH transition.}
   \label{Table-of-exponents}
\vskip -5mm
\end{table}

The class C network model can be driven away from its (multi)critical point given by Eq. (\ref{critical-point}) (and $\delta = 0$) in different ways. Taking $t_{A\sigma}\neq t_{B\sigma}$ (but keeping $t_{S\up} = t_{S\down}$) is the only perturbation that preserves the class C symmetries. It drives the system through a SQH transition between an insulator and a SQH state. Introducing a uniform Zeeman field (or a non-zero chemical potential) breaks the global spin-rotation symmetry, and splits the transition into two ordinary IQH transitions, each in class A. The same effect is achieved by making $t_{S\up} \ne t_{S\down}$.

To describe these relevant perturbations in a quantitative way, let us parametrize the node scattering amplitudes in the vicinity of the critical point (\ref{critical-point}) as follows:
\begin{align}
t_{A\up} &= t(1 + \epsilon + \Delta), & t_{A\dn} &= t(1 + \epsilon - \Delta), \nonumber \\
t_{B\up} &= t(1 - \epsilon + \Delta), & t_{B\dn} &= t(1 - \epsilon - \Delta).
\label{parameters}
\end{align}
Then nonzero $\epsilon$, Zeeman field $B_z$ (or $\eta$), and $\Delta$ are all relevant perturbations that induce finite localization lengths scaling as
\begin{align}
\xi &\sim |\epsilon|^{-\nu}, & \xi_B &\sim |\eta|^{-\nu_B}, & \xi_\Delta &\sim |\Delta|^{-\mu_\Delta}.
\end{align}
Thus defined critical exponents have been analytically determined in Ref. [\onlinecite{GruzReadLudwig:99}], where the authors suggested that the parameter $\Delta$ may describe a {\it random} Zeeman field. The exponents were numerically studied in Refs. [\onlinecite{Kagalovsky:99, SenthilMarstonFisher:99, Kagalovsky:2001, Kagalovsky:Annalen, Obuse-et-al:10}]. The results are summarized in the first three columns of Table \ref{Table-of-exponents}.

A microscopically distinct perturbation of the class C network that induces a crossover to class A is the introduction of a nonzero phase $\delta$ in the link matrices (\ref{link-matrix-U2}). The case of a random extra phase $\delta$ with zero mean and variance $p^2$ was numerically studied in Ref. [\onlinecite{Kagalovsky:Annalen}], and that of a constant phase $\delta_0$ --- in Ref. [\onlinecite{Kagalovsky:2001}]. Both perturbations appeared to be relevant, as expected on symmetry grounds, and resulted in finite localization lengths  that scaled as
\begin{align}
\xi_p &\sim p^{-\mu_p}, & \xi_{\delta_0} &\sim |\delta_0|^{-\mu_{\delta_0}}.
\end{align}

From our comments above it should be clear that a constant phase $\delta_0$ is exactly equivalent to a uniform nonzero Zeeman field, which immediately implies
\begin{align}
\mu_{\delta_0} = \nu_B.
\end{align}
By the same token, a random phase $\delta$ is equivalent to a random Zeeman field. In the rest of the paper we assume that the phases $\delta$ on the links are independent identically distributed random variables with the mean $\delta_0$ and variance $p^2$, where both quantities are small: $\delta_0, p \ll 1$.  In the following sections we will show that the small parameters $\eta$, $\delta_0$, and $p$ always appear in the combination
\begin{align}
\lambda \equiv \eta - i\delta_0 + p^2.
\label{lambda}
\end{align}
This immediately implies that
\begin{align}
\mu_p = 2 \mu_{\delta_0} = 2\nu_B.
\label{lambda-exponents}
\end{align}

In subsequent sections we will use the SUSY method and the mapping to percolation to obtain the exact values of the exponents $\mu_{\delta_0}$ and, therefore, $\mu_p$. In addition, our analysis uncovers a subtle mistake made in Ref. [\onlinecite{GruzReadLudwig:99}] that led to a wrong prediction for the exponent $\mu_\Delta$. After correcting the mistake, we obtain the values of the exponents shown in Table \ref{Table-of-exponents}. We also show in the Table a numerical value for the exponent $\mu_p$ obtained by a direct computer simulation of the class C network model with an additional random phase $\delta$ on the links.

\section{Supersymmetric transfer matrices}
\label{sec:SUSY}

\begin{figure}[t*]
\centering \includegraphics[width=\columnwidth]{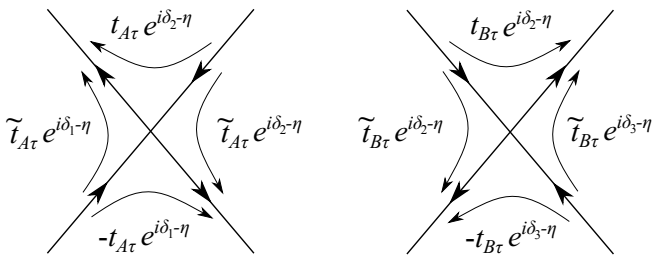}
\caption{The scattering amplitudes at the nodes on the two sublattices. The node scattering matrices are diagonal in the spin indices, so we only show one channel per link.}
\label{Fig:Node}
\vskip -3mm
\end{figure}

In this section we apply the SUSY method \cite{Read, GRS:97} to our network model in the vicinity of its critical point to map it to classical percolation. Before we describe the technical steps, let us present our strategy. As we mentioned in the previous section, the mapping to percolation is only possible when the scattering on the links is described by SU(2) matrices. Thus, a direct application of this method to our system, where the link matrices are given by Eq. (\ref{link-matrix-eta}), is impossible. We circumvent this difficulty as follows. We can always remove the factors $e^{i\phi} = e^{i\delta - \eta}$ from the link scattering matrices and reassign them to the nodal matrices in such a way that the Green's function (\ref{Greens-function}) is not affected. The redefined nodal matrices become
\begin{align}
{\cal S}_{S\sigma} = \begin{pmatrix} r_{S\sigma} e^{i\phi} &  t_{S\sigma}^{\vphantom{2}} e^{i\phi'} \\
- t_{S\sigma}^{\vphantom{2}} e^{i\phi} &  r_{S\sigma} e^{i\phi'} \end{pmatrix}.
\label{node-matrices-new}
\end{align}
Here $\delta$ and $\delta'$ in $\phi$ and $\phi'$ are independent, since they come from two different links incoming at a node, see Fig. \ref{Fig:Node}. Having shifted the factors $e^{i\phi}$ onto the nodes, we are now free to perform the SU(2) average on the links. Subsequently, we can perform the average over the phases $\delta$.

In the SUSY method the vertical direction in Fig. \ref{Fig:Network} is regarded as the (imaginary) time $\tau$. The vertical zig-zags of links that go up (along the time direction) correspond to sites of a quantum one-dimensional chain with an odd label $i$. The down-going links correspond to even sites. At each odd site $i$ there is a Fock space ${\cal F}_i = {\cal F}_{i \up} \otimes {\cal F}_{i \down}$ of fermions and bosons, and at each even site the Fock space is $\bar{\cal F}_i = \bar{\cal F}_{i \up} \otimes \bar{\cal F}_{i \down}$. The spaces on the odd and even sites differ by the commutation relations for creation and annihilation operators of the up and down particles, see Appendix \ref{app:SUSY-class-C}.

Scattering of fluxes on links of the network is represented by the second-quantized transfer matrices $T_{2i-1}$ and $T_{2i}$ which describe the evolution of states in ${\cal F}_{2i-1}$ or $\bar{\cal F}_{2i}$ between two discrete imaginary time slices through the lower and upper half-link. Scattering at a node on sublattice $A$ is represented by the transfer matrix $T_{2i-1,2i}$ which evolves states in the tensor product ${\cal F}_{2i-1} \otimes \bar{\cal F}_{2i}$ between two discrete imaginary time slices (below and above the node), and similarly for the $B$ sublattice. All second-quantized transfer matrices are exponentials of quadratic forms in creation and annihilation operators, see details in Appendix A.

As is known from Ref. [\onlinecite{GruzReadLudwig:99}] (and reviewed in Appendix \ref{app:SUSY-class-C}), in the spin-rotation invariant case ($t_{S\up} = t_{S\down}$ and $\delta = 0$), the transfer matrices commute with the sum over sites of the eight generators (superspin components) of the superalgebra $\mathfrak{osp}(2|2) \cong \mathfrak{sl}(2|1)$. The eight generators of $\mathfrak{osp}(2|2)$ on each site appear as all bilinears in the fermions and bosons and their adjoints, which are singlets under the random SU(2). These are denoted by \cite{snr} $B$, $Q_3$, $Q_\pm$, $V_\pm$, $W_\pm$ for the up sites (and with bars for the down sites) and have similar expressions for the two types of sites. We combine the generators on each site into a single eight-component object, a superspin, and call it $J_{2i-1}$ and ${\bar J}_{2i}$ for up and down sites. Breaking the spin-rotation invariance by either of the symmetry-breaking perturbations, breaks the SUSY of the transfer matrices down to $\mathfrak{gl}(1|1)$ generated on each up site by $K = \{B, Q_3, V_-, W_+\}$ (similarly for the down sites).

Averaging over the random SU(2) matrices on the links projects each Fock space ${\cal F}_{2i-1}$ ($\bar{\cal F}_{2i}$) onto a three-dimensional subspace which is the fundamental (dual to the fundamental) representation of  $\mathfrak{osp}(2|2)$. In the notation of Ref. [\onlinecite{fss}] these irreducible representations (irreps) are $\pi(\pm\tfrac{1}{2}, \tfrac{1}{2})$. For a single transfer matrix the projection (see Appendix \ref{app:node-T} for details) results in
\begin{widetext}
\begin{align}
{\hat P} T {\hat P}  &= 1 + \big(r_\up r_\down e^{2i\phi} - 1\big) (B + Q_3) - \big(r_\up r_\down e^{2i{\bar \phi}} - 1\big)(\Bb + \Qb_3)
- \big(r_\up r_\down e^{2i\phi} - 1\big)\big(r_\up r_\down e^{2i{\bar \phi}} - 1\big) (B + Q_3)(\Bb + \Qb_3)  \nonumber \\
& \quad + e^{2i(\bar{\phi} + \phi)} \Big[t_\up^2 t_\down^2 (B + Q_3)(\Bb + \Qb_3) - \frac{t_\up^2 + t_\down^2}{2} (K \cdot \Kb - B - Q_3 + \Bb + \Qb_3) \Big] \nonumber \\
& \quad - t_\up t_\down e^{2i\bar{\phi}}(Q_+ \Qb_- + V_+ \Wb_-) - t_\up t_\down e^{2 i \phi}(Q_- \Qb_+ - W_- \Vb_+).
\label{withRandomPhase}
\end{align}
\end{widetext}
Here we have suppressed the sublattice index for brevity, and also used the $\mathfrak{gl}(1|1)$-invariant product of the superspins $K$ and $\Kb$:
\begin{align}
K \cdot \Kb &= 2Q_3 \bar{Q}_3 - 2B\bar{B}  -  V_- \bar{W}_+ + W_+ \bar{V}_-.
\end{align}

We can now carry out averages (that we denote by angular brackets) over the independent random phases $\delta$. Using the notation
\begin{align}
\langle e^{2i\phi} \rangle = \langle e^{2i{\bar \phi}} \rangle \equiv \Lambda,
\end{align}
and the $\mathfrak{osp}(2|2)$-invariant product
\begin{align}
J\cdot \Jb &= 2Q_3 \bar{Q}_3 - 2B\bar{B}  -  V_- \bar{W}_+ + W_+ \bar{V}_- \nonumber \\
&\quad + V_+ \bar{W}_- - W_- \bar{V}_+ + Q_+ \bar{Q}_- + Q_-\bar{Q}_+,
\end{align}
we can write the average of the projected transfer matrix on the sublattice $S$ as
\begin{align}
& \langle {\hat P} T_S {\hat P}\rangle = 1 + c_{S1} J \cdot \Jb + c_{S2} K \cdot \Kb \nonumber \\
& + c_{S3} (B + Q_3)(\Bb + \Qb_3) + c_{S4} (B + Q_3 - \Bb -\Qb_3),
\label{PTP-exact}
\end{align}
where the coefficients are given by
\begin{align}
c_{S1} &= - \Lambda t_{S\up} t_{S\down}, \nonumber \\
c_{S2} &= - \frac{1}{2} \Lambda^2(t_{S\up}^2 + t_{S\down}^2) + \Lambda t_{S\up} t_{S\down}, \nonumber \\
c_{S3} &= \Lambda^2 t_{S\up}^2 t_{S\down}^2 - (\Lambda r_{S\up} r_{S\down} - 1)^2, \nonumber \\
c_{S4} &= \frac{1}{2} \Lambda^2(t_{S\up}^2 + t_{S\down}^2) + \Lambda r_{S\up} r_{S\down} - 1.
\label{coefficients}
\end{align}

When the extra phases $\phi$ vanish ($\Lambda = 1$), the coefficients in this expression simplify, and it reduces to the one studied before in Ref. [\onlinecite{GruzReadLudwig:99}]:
\begin{align}
& \langle {\hat P} T_S {\hat P}\rangle = 1 - t_{S\up} t_{S\down} J \cdot \Jb - \frac{(t_{S\up} - t_{S\down})^2}{2} K \cdot \Kb
\nonumber \\ &
- \frac{(r_{S\up} - r_{S\down})^2}{2} \big[2 (B + Q_3)(\Bb + \Qb_3) + B + Q_3 - \Bb -\Qb_3 \big].
\end{align}
In particular, if no class C symmetries are broken ($t_{S\up} = t_{S\down}$), the average transfer matrix reduces to
\begin{align}
\langle {\hat P} T_S {\hat P}\rangle = 1 - t_{S}^2 J \cdot \Jb,
\label{PTP-class-C}
\end{align}
an expression that was interpreted in terms of classical bond percolation on a square lattice in Ref. [\onlinecite{GruzReadLudwig:99}].

\section{Superspin chain and critical exponents}
\label{sec:superspin-chain}

So far everything was exact. Now we will perform an additional step that is useful in the study of network models. This is to consider an anisotropic limit, when all amplitudes $t_{S\sigma}$ are small [which can be achieved by taking $t \ll 1$ in Eq. (\ref{parameters})]. In this case the disorder-averaged product of all 
transfer matrices can be written as an evolution operator in continuous imaginary time $\tau$:
\begin{align}
U = \exp \Big( - \int \!\! d\tau \, {\cal H}_{1\text{D}} \Big),
\label{contevolution}
\end{align}
where the effective Hamiltonian ${\cal H}_{1\text{D}}$ describes a 1D superspin chain, with alternating $\pi(\pm\tfrac{1}{2}, \tfrac{1}{2})$  representations (superspins) on each site along the chain. The spin chain has a critical point, and various deviations from it appear at this step as perturbations of the critical Hamiltonian.

When passing to the anisotropic limit, we will expand the coefficients in expressions for average transfer matrices to leading order in all small parameters ($t, \epsilon, \Delta, \eta, \delta_0, p$).

First, consider the maximally symmetric case (\ref{PTP-class-C}):
\begin{align}
\langle {\hat P} T_{2i-1,2i} {\hat P}\rangle &\approx 1 - t^2(1 + 2\epsilon) J_{2i-1} \cdot \Jb_{2i}, \nonumber \\
\langle {\hat P} T_{2i,2i+1} {\hat P}\rangle &\approx 1 - t^2(1 - 2\epsilon) \Jb_{2i} \cdot J_{2i+1}.
\end{align}
Combining all transfer matrices, we obtain the effective 1D Hamiltonian
\begin{align}
{\cal H}_{1\text{D}} = {\cal H}_0 + {\cal H}_1,
\end{align}
where
\begin{align}
{\cal H}_0 = t^2 \sum_i \big(J_{2i-1} \cdot \Jb_{2i} + \Jb_{2i} \cdot J_{2i+1} \big)
\end{align}
describes the critical superspin chain, and the staggered term
\begin{align}
{\cal H}_1 &= 2 t^2 \epsilon \sum_i D_i, & D_i &= J_{2i-1} \cdot \Jb_{2i} - \Jb_{2i} \cdot J_{2i+1},
\end{align}
represents a relevant perturbation. As was argued in Ref. [\onlinecite{GruzReadLudwig:99}], the dimer operator $D_i$ represents the {\it two-hull} operator in the critical percolation picture, with dimension $x_2 = 5/4$. The corresponding critical exponent is
\begin{align}
\nu = (2 - x_2)^{-1} = \frac{4}{3}.
\end{align}

Having identified the role of the sublattice asymmetry $\epsilon$, let us turn to the general case, Eqs. (\ref{PTP-exact}) and (\ref{coefficients}). The terms with the coefficients $c_{S2}$ and $c_{S3}$ contain bilinears in the superspins, and can be thought of as introducing two kinds of anisotropy in the superspin space. The last term (with the coefficient $c_{S4}$ is linear in superspins, ans corresponds to the {\it one-hull} operator in the critical percolation picture, with dimension $x_1 = 1/4$. This is the lowest among the dimensions of operators in critical percolation. Thus, even without the knowledge of the dimensions of the anisotropic terms, we can claim that the last term in Eq. (\ref{PTP-exact}) is the most relevant perturbation. We will see that all symmetry-breaking perturbations couple to this term, and it, therefore, determines the corresponding critical exponents.

In the anisotropic limit and close to the critical point we have, first of all
\begin{align}
\Lambda &\approx 1 - 2\lambda,
\end{align}
where $\lambda$ is given in Eq. (\ref{lambda}). Expanding the coefficients $c_{S1}$ in Eq. (\ref{PTP-exact}) we have
\begin{align}
c_{A1} &\approx -t^2 (1 + 2 \epsilon - \Delta^2 - 2\lambda), \nonumber \\
c_{B1} &\approx -t^2 (1 - 2 \epsilon - \Delta^2 - 2\lambda).
\end{align}
We see that the symmetry-breaking perturbations simply renormalize the coupling constant $t^2$ of the critical Hamiltonian ${\cal H}_0$.

In the other terms it is sufficient to set $\epsilon = 0$. Then the coefficients on the two sublattices coincide, and their expansions look like
\begin{align}
c_{S2} & \approx - 2 t^2 (\Delta^2 - \lambda), \nonumber \\
c_{S3} & \approx - 4 t^4 \Delta^2 - 4 t^2 \lambda, \nonumber \\
c_{S4} & \approx -2 t^4 \Delta^2 - 2 (1 + t^2) \lambda.
\end{align}
The last expression confirms the conclusion of Ref. [\onlinecite{GruzReadLudwig:99}] that the non-zero energy $\eta$ couples to the most relevant perturbation, the one-hull operator $B + Q_3 - \Bb -\Qb_3$ (that happens to represent the local density of states), and leads to a localization length that has a power-law behavior with the exponent
\begin{align}
\nu_B = (2 - x_1)^{-1} = \frac{4}{7}.
\end{align}
Moreover, since $\eta$ enters all expressions in the combination $\lambda = \eta - i\delta_0 + p^2$, we immediately obtain the equalities between critical exponents given in Eq. (\ref{lambda-exponents}).

Next we see that the {\it square} of the spin-rotation symmetry-breaking parameter $\Delta$ also couples to the one-hull operator. This immediately implies that
\begin{align}
\mu_\Delta = \mu_p = \frac{8}{7}.
\end{align}
This value is different from the result $\mu_\Delta = 3/2$ obtained in Ref. [\onlinecite{GruzReadLudwig:99}]. The reason for this difference is that $\Delta^2$ enters the coefficient of the one-hull term $B + Q_3 - \Bb - \Qb_3$ in the combination $t^4 \Delta^2$. In taking the anisotropic limit $t \to 0$ this term was neglected in Ref. [\onlinecite{GruzReadLudwig:99}]. Instead, in that reference the authors argued that the exponent $\mu_\Delta$ is determined by the dimension of the superspin anisotropy operator $K \cdot \Kb$, which was conjecturally found (and was larger than $x_1 = 1/4$). However, since the one-hull term is the most relevant scaling operator, this term really determines the scaling behavior of the localization length $\xi_\Delta$ for any finite $t$.

\section{Numerical Results}
\label{sec:numerics}

In this section we compare the exact values of critical exponents obtained above with results of numerical simulations.

First we report our numerical results for the exponent $\mu_p$. We have simulated the SQH network model with only one relevant perturbation: extra random phases on the links with the  mean $\delta_0 = 0$ and variance $p$. The other perturbations ($\epsilon$, $\eta$, $\Delta$) were set to zero. We used the standard transfer matrix method in the quasi-one-dimensional geometry with periodic boundary conditions in the transverse direction (cylinder) \cite{TM-method-1, TM-method-2}. Our system lengths reached $10^{6}$, and the circumferences $M$ ranged from 32 to 192 with various random phase variances $p$.

Without symmetry-breaking perturbations, all Lyapunov exponents of the transfer matrix product are doubly degenerate due to the presence of time-reversal invariance (Kramers degeneracy). It was suggested in Ref. [\onlinecite{Kagalovsky:99}] that when the time-reversal symmetry is broken by a small perturbation, the renormalized localization length (the inverse of the smallest positive Lyapunov exponent) and  the deviation from Kramers degeneracy $\bar \xi$ (the difference between the two smallest positive Lyapunov exponents multiplied by the circumference $M$) exhibit scaling behavior characterized by the same exponent. This idea was further supported in Refs. [\onlinecite{Kagalovsky:Annalen}] and [\onlinecite{Kagalovsky:2001}]. It turns out that the deviation from Kramers degeneracy $\bar \xi$ is a superior way to extract critical exponents in this case, since we know its exact value $\bar \xi$ = 0 at the critical point.

\begin{figure}[t]
\vskip -5mm
\centering \includegraphics[width = \columnwidth]{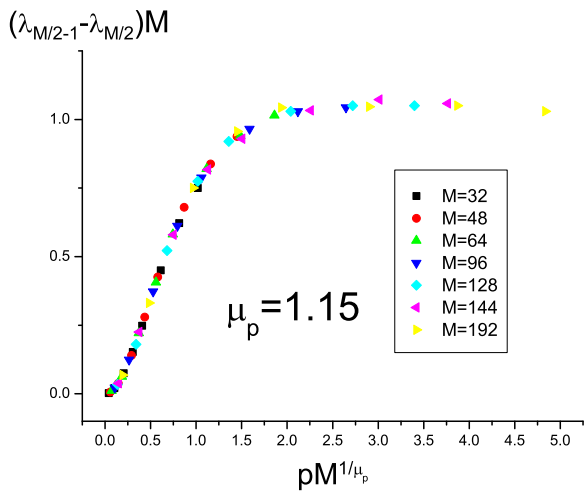}
\vskip -5mm
\caption{(Color online) Deviation from Kramer's degeneracy as function of $p M^{1/\mu_p}$ with $\mu_p=1.15$ for $\epsilon=\Delta=0$.}
\label{p-mu-fig}
\vskip -3mm
\end{figure}

Thus, in Fig. \ref{p-mu-fig} we present a one-parameter scaling results for ${\bar \xi} \equiv (\lambda_{M/2-1} - \lambda_{M/2)})M$ as a function of the scaling variable $x \equiv p M^{1/\mu_p}$: ${\bar \xi} = f(x)$. In order to improve the accuracy we do not use data for systems with small circumferences, and obtain the value of the critical exponent using an optimization program that ensures the best scaling collapse. The routine determines the least-squares approximation to the scaling function $f(x)$ in terms of the Chebyshev polynomials by minimizing the sum of squares of the deviations of the data points from the corresponding values of the polynomial, while also varying the exponent $\mu_p$. The result $\mu_p = 1.15$ is in excellent agrement with our analytical prediction $\mu_p = 8/7 \approx 1.14$.


Now we comment on previously published numerical results for various critical exponents. In this paper we find a perfect agreement between the analytical prediction and the numerical value of $\mu_p$. As we have shown above, the values of $\mu_p$ and $\mu_\Delta$ must be the same. On the other hand, a numerical result found in Ref. [\onlinecite{Kagalovsky:99}] was $\mu_\Delta \approx 1.45$.  We believe that the reason for this discrepancy is that only large values of $\Delta$ were used in Ref. [\onlinecite{Kagalovsky:99}]. Indeed,  in that paper it was impossible to resolve two separate critical states for $\Delta \leqslant 0.5$.

A similar discrepancy exist between the numerical values of the exponent $\nu$ reported in different papers. In the original paper \cite{Kagalovsky:99} a broad range of $\epsilon \in [0,1]$ was used (including the values of $\epsilon$ far from the critical point), and the result was $\nu \approx 1.12$. In a more recent study \cite{Obuse-et-al:10} the authors used only data for $\epsilon < 0.05$ (very close to the critical point), and obtained $\nu \approx 1.335$, in excellent agreement with the analytical prediction $\nu = 4/3$. The same arguments explain the discrepancy between the exact value $\mu_B = \mu_{\delta_0} = 4/7 \approx 0.57$ and the numerical result $\mu_{\delta_0} \approx 0.7$ \cite{Kagalovsky:2001}. Convincing arguments for the necessity to use only the data very close to the critical point for accurate results on critical phenomena are presented in Ref. [\onlinecite{Evers_Gruzberg_Hideaki:2012}].

\section{Conclusions}

In conclusion, we have studied relevant perturbations at the spin quantum Hall (SQH) transition critical point. Many critical exponents at the transition have been found before. We have derived several new exponents, and corrected a subtle error in an earlier prediction. All (present and older) results are summarized in Table \ref{Table-of-exponents}. Our analysis demonstrates that several symmetry-breaking perturbations, which are distinct in the microscopic network model of the SQH transition, correspond to the same relevant perturbation at the critical point. In particular, the variance $p$ of the extra random phase of the scattering matrices on the links plays exactly the same role as the spin-rotation symmetry-breaking parameter $\Delta$. Both happen to represent the effect of a random Zeeman magnetic field and drive the system to a localized phase.

\begin{figure}[t]
\centering \includegraphics[width = \columnwidth]{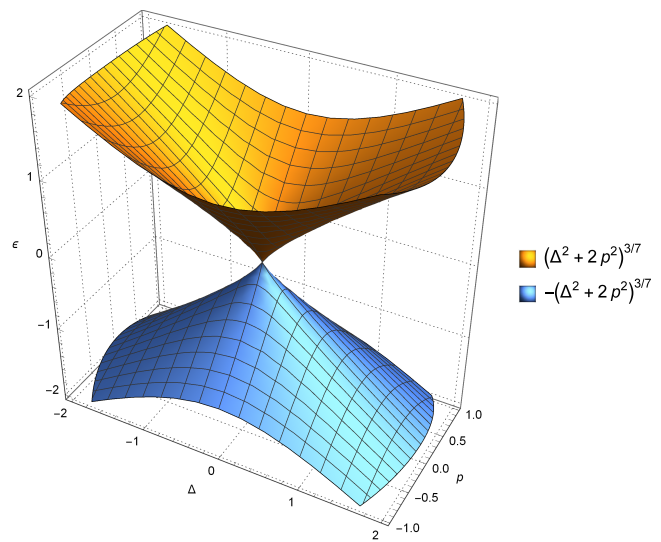}
\caption{(Color online) Schematic plot of the phase boundary $\epsilon = \pm(a \Delta^2 + b p^2)^{\varphi/2}$.}
\label{fig:phase-diagram}
\vskip -3mm
\end{figure}

Our results allow us to represent the phase diagram for our system in the three-dimensional space of parameters $\epsilon$, $\Delta$, and $p$. Indeed, the last two parameters appear in the combination $a \Delta^2 + b p^2$ with some non-universal coefficients. By the standard scaling argument, the critical surface is described by the equation
\begin{align}
\epsilon = \pm(a \Delta^2 + b p^2)^{\varphi/2},
\label{critical-surface}
\end{align}
where the crossover exponent
\begin{align}
\varphi = \mu_p/\nu = 6/7.
\end{align}
The critical surface is schematically shown in Fig. \ref{fig:phase-diagram}, where, for illustration purposes, we chose $a=1$, $b=2$.

Finally, it is interesting to note that when we set $\epsilon = 0$ in our model, then for any nonzero $p$ it becomes formally equivalent to the network model proposed in Refs. [\onlinecite{LeeChalker}] as a tool to study localization of electrons in a random magnetic field. While the physics of the random magnetic field problem and the spin quantum Hall effect is very different, the equivalence of the models is seen, in particular, in the absence of extended states in the $\Delta$-$p$ plane except for the critical point at the origin.

\section*{Acknowledgements}
We thank Mikhail Raikh for many fruitful discussions. This work was supported in part by US - Israel Binational
Science Foundation (BSF) Grant No. 2010030 and by the Shamoon College of Engineering (SCE) under internal Grant No. 5368911113.

\appendix

\section{SUSY method for the SU(2) network in class C}
\label{app:SUSY-class-C}

In this Appendix we provide details of the SUSY method for the class C network.

Usually in the SUSY approach one needs two types of bosons and fermions, retarded and advanced, to be able to obtain two-particle properties. However, the particle-hole symmetry relates retarded and advanced Green’s functions \cite{SenthilMarstonFisher:99}. Hence, for the study of mean values of simple observables, we need only one fermion and one boson per spin direction per site. Let us now consider transfer matrices on the up- and down-going links separately.

\subsection{Up-links}

Let us denote the single boson and fermion per spin direction $\sigma$ for an up site $i$ as $f_{i\sigma}, b_{i\sigma}$. In our scheme of labeling the up sites have an odd index $i$. Propagation of a doublet of complex fluxes on a link is governed by an SU(2) scattering matrix ${\cal S}_0$, which relates the doublets of incoming $I$ and outgoing $O$ fluxes:
\begin{align}
\label{smat}
\begin{pmatrix} o_\up \\ o_\down \end{pmatrix} = {\cal S}_0 \begin{pmatrix} i_\up \\ i_\down \end{pmatrix}
= \begin{pmatrix} \alpha & \beta \\ -\beta^* & \alpha^* \end{pmatrix}
\begin{pmatrix} i_\up \\ i_\down \end{pmatrix}.
\end{align}
This propagation looks identical to the propagation of fluxes on two adjacent links of a directed network which was considered in detail in Ref. [\onlinecite{GRS:97}]. Thus, we can easily borrow the second-quantized supersymmetric form of the transfer matrix from that reference by omitting the advanced particles and replacing the site indices by the spin indices:
\begin{align}
T &= \, :\!\exp \biggl({\beta \over \alpha^*} (\fdup \fdown + \bdup \bdown) -
{\be^* \over \al} (\fddown \fup + \bddown \bup) \biggr)\!: \no\\
& \quad \times \al^{n_{f\up}+n_{b\up}} (\al^*)^{n_{f\down} + n_{b\down}},
\end{align}
where $n_{b\sigma} = \bd_\sigma \bph_\sigma$, etc., and the colons stand for normal ordering.

It is easy to obtain the commutation relations between $T$ and fermions and bosons:
\begin{align}
\label{commrels}
T \cdup &= (\alpha \cdup - \be^* \cddown) T, & T \cddown & = (\al^* \cddown + \beta \cdup) T, \nonumber \\
T \cup &= (\al^* \cup - \beta \cdown) T, & T \cdown &= (\alpha \cdown + \be^* \cup) T.
\end{align}
Here and later by $c, \cd$ we denote either $b$ or $f$ and their conjugates. These relation are conveniently interpreted as giving the evolution of states created by $\cd_\sigma$ in the Schr\"odinger representation, or the operators $\cd_\sigma$ themselves in the Heisenberg representation, where the operators are time ordered from right to left. Equations (\ref{commrels}) may be written in a short form as
\begin{align}
\label{comrel}
T \cd_\sigma &= \cd_{\sigma'} {\cal S}^{\vphantom{\dagger}}_{\sigma'\sigma} T,
& T \cph_{\sigma} &= {\cal S}^\dagger_{\sigma \sigma'} \cph_{\sigma'} T.
\end{align}

\begin{figure}
\includegraphics[width=1.8in]{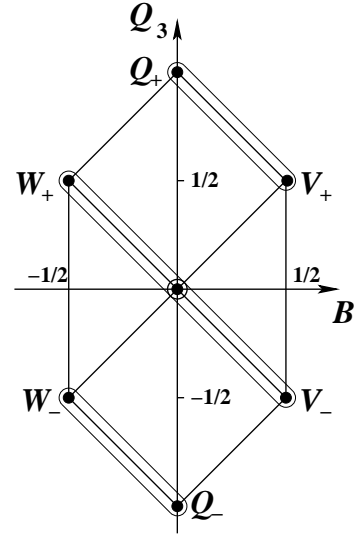}
\vskip 10mm
\caption{The weights of the adjoint representation of $\mathfrak{osp}(2|2)$. We show two doublets and the adjoint of the subalgebra gl$(1|1)$.}
\label{weights-adjoint}
\end{figure}

Relations (\ref{comrel}) imply that under the commutation with $T$ the bosons and fermions transform as spinors (in the fundamental representation) of the SU(2) group of the scattering matrices ${\cal S}_0$. Then the SU(2) singlet bilinear combinations of our fermionic and bosonic operators commute with the $T$. There are 8 such combinations, which we denote following Ref. [\onlinecite{snr}] as
\begin{align}
\label{up-gen}
B &= \frac{1}{2}(\bdup \bup + \bddown \bdown + 1), & Q_3 &= \frac{1}{2}(\fdup \fup + \fddown \fdown - 1), \no \\
Q_+ &= \fdup \fddown, & Q_- &= \fdown \fup, \no \\
V_+ &= \frac{1}{\sqrt 2} (\bdup \fddown - \bddown \fdup), & W_- &= (V_+)^\dagger, \no \\
V_- &= - \frac{1}{\sqrt 2} (\bdup \fup + \bddown \fdown), & W_+ &= - (V_-)^\dagger.
\end{align}
We combine these generators into a single eight-component object $J$, or superspin. These operators satisfy the (anti)commutation relations of the $\mathfrak{osp}(2|2)$ Lie superalgebra:
\begin{gather}
[B,Q_3] = [B,Q_\pm] = 0, \nonumber \\
\begin{align}
[B,V_\pm] &= \frac{1}{2} V_\pm, & [B,W_\pm] &= -\frac{1}{2} W_\pm, \nonumber \\
[Q_3, Q_\pm] &= \pm Q_\pm, & [Q_+, Q_-] &= 2Q_3, \nonumber \\
[Q_3, V_\pm] &= \pm \frac{1}{2}V_\pm,
& [Q_3, W_\pm] &= \pm\frac{1}{2}W_\pm, \nonumber \\
[Q_+, V_-] &= V_+, & [Q_+, W_-] &= W_+, \nonumber \\
[Q_-, V_+] &= V_-, & [Q_-, W_+] &= W_-, \nonumber
\end{align} \\
[Q_+, V_+] = [Q_+, W_+ ] = [Q_-, V_-] = [Q_-,W_- ] = 0, \nonumber \\
\{V_+,V_-\} = \{W_+,W_-\} = 0, \nonumber \\
\{V_+, W_+\} = Q_+, \quad\,\,\, \{V_+, W_-\} = B-Q_3, \nonumber \\
\{V_-, W_-\} = -Q_-, \quad \{V_-, W_+\} = -B - Q_3.
\label{comm-rels}
\end{gather}
The components $B$ and $Q_3, Q_\pm$ of the superspin generate the even subalgebra $\mathfrak{u}(1) \oplus \mathfrak{su}(2)$. An important sub-superalgebra is the $\mathfrak{gl}(1|1)$ formed by $Q_3, B, V_-, W_+$, which we will call collectively the components of the superspin $K$.

The algebra $\mathfrak{osp}(2|2)$ has rank two (it has two Cartan generators: $B$ and $Q_3$), and its representations are labeled by two quantities. We use the u(1) ``charge'' (the value of $B$ in a representation) $b$, and the value $q$ of the ``spin'' of su(2) generated by the $Q_i$. Representation with the highest weight $(b,q)$ is denoted by $\pi(b,q)$ \cite{fss}. For example, the adjoint representation of $\mathfrak{osp}(2|2)$ is $\pi(0,1)$, and it is shown in Fig. \ref{weights-adjoint}.

The quadratic Casimir of $\mathfrak{osp}(2|2)$ is
\begin{align}
\label{casimir-J}
C_2(J) &=  Q_3^2  - B^2 + \frac{1}{2}\big( Q_-Q_+ + Q_+Q_- + V_+W_- \no \\
&\quad -W_-V_+ - V_-W_+ +W_+V_- \big),
\end{align}
and in the representation $\pi(b,q)$ it takes the value $q^2 - b^2$. The quadratic Casimir of the $\mathfrak{gl}(1|1)$ subalgebra is 
\begin{align}
\label{casimir-K}
 C_2(K) &=  Q_3^2  - B^2 + \frac{1}{2}\big(W_+V_- - V_-W_+ \big)
\end{align}

\begin{figure}
\includegraphics[width=3in]{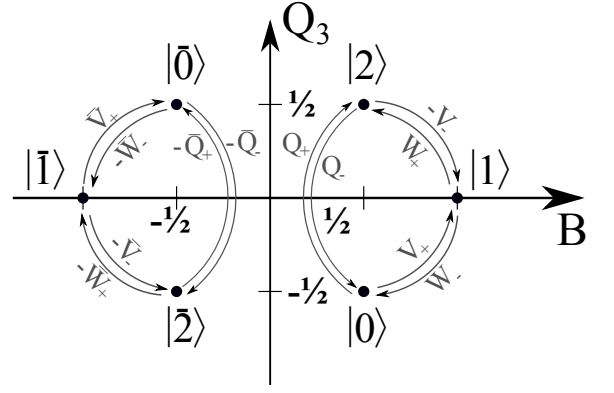}
\vskip 5mm \caption{The weights of the fundamental representation $\pi(\tfrac{1}{2},\tfrac{1}{2})$ of $\mathfrak{osp}(2|2)$ and its dual $\pi(-\tfrac{1}{2},\tfrac{1}{2})$.}
\label{weights-fundamental}
\end{figure}

As follows from Eqs. (\ref{comrel}) the action of $T$ decomposes the Fock space of the bosons and fermions  into irreducible representations of SU(2). When we average over the SU(2), any non-trivial representation is projected out. Thus, the averaged link transfer matrix acts as the projection operator to the subspace of the SU(2)-singlets:
\begin{align}
\la T \ra_{\text{SU(2)}} = P.
\label{proj}
\end{align}
There are only three singlets in this subspace which we denote as $|m\ra$, $m = 0, 1, 2$, and define as
\begin{align}
|0\ra &= |\text{vacuum}\ra,  \\
|1\ra &= V_+|0\ra = \frac{1}{\sqrt 2}(\bdup \fddown - \bddown \fdup)|0\ra, \\
|2\ra &= Q_+|0\ra = \fdup\fddown |0\ra.
\end{align}
These singlets form the fundamental representation $\pi(\tfrac{1}{2},\tfrac{1}{2})$ of the $\mathfrak{osp}(2|2)$ algebra. It is shown in Fig. \ref{weights-fundamental} together with its dual $\pi(-\tfrac{1}{2},\tfrac{1}{2})$.

It is easy to find the action of the generators of $\mathfrak{osp}(2|2)$ on the states in the fundamental representation:
\begin{align}
Q_3 |0\ra &=  -\frac{1}{2} |0\ra, & Q_3 |1\ra &= 0, & Q_3 |2\ra &= \frac{1}{2} |2\ra, \nonumber \\
B |0\ra &= \frac{1}{2} |0\ra, & B |1\ra &= |1\ra, & B |2\ra &= \frac{1}{2} |2\ra, \nonumber \\
Q_+ |0\ra  &= |2\ra, & Q_- |2\ra &= |0\ra, \nonumber \\
V_+ |0\ra &= |1\ra, & V_- |2\ra &= -|1\ra, \nonumber \\
W_+ |1\ra &= |2\ra, & W_- |1\ra &= |0\ra.
\label{J-action}
\end{align}

\subsection{Down-link}

Consider next a down-going link. Fermions and bosons on such links will be denoted by bars: $\fb_\sigma, \bb_\sigma$. On a down link the incoming and outgoing channels are interchanged. Then if we want to think of the evolution of the states on the link as going up in the vertical time direction, we need to relate the doublet $I$ to the doublet $O$. Inverting relations (\ref{smat}), we get
\begin{align}
\label{sdag}
\begin{pmatrix} i_\up \\ i_\down \end{pmatrix} = {\cal S}_{\rm link}^\dagger \begin{pmatrix} o_\up \\ o_\down \end{pmatrix}
= \begin{pmatrix} \al^* & -\beta \\ \be^* & \alpha \end{pmatrix} \begin{pmatrix} o_\up \\ o_\down \end{pmatrix}.
\end{align}
Then the bosonic part of the transfer matrix on a down link is
\begin{align}
\label{Tbbar}
{\bar T}_{\bb} = \,:\exp \biggl(-{\beta \over \al}
\bbdup \bbdown + {\be^* \over \al^*} \bbddown \bbup \biggr)\!:
(\al^*)^{n_{\bb\up}} \al^{n_{\bb\down}}.
\end{align}

This transfer matrix gives the following commutation relations for the bosons:
\begin{align}
\label{comrelbar}
{\bar T}_{\bb} \bbd_\sigma &= \bbd_{\sigma'} {\cal S}^\dagger_{\sigma'\sigma} {\bar T}_{\bb}, &
{\bar T}_{\bb} \bb_\sigma &= {\cal S}_{\sigma\sigma'} \bb_{\sigma'} {\bar T}_{\bb}.
\end{align}
These relations are again easily interpreted from the point of view of evolution of states. Comparing them with Eq. (\ref{comrel}), we see that on the down links the states at later times are related to the states at earlier times in the opposite way to what happens on the up links, which is natural.

Now we want to add fermions. This is somewhat tricky, since the cancellation of closed loops in the SUSY formalism requires the presence of negative norm states in the same way as in the case of the Chalker-Coddington model. So far we used the canonical bosons $\bb$. Then the fermions on the down links should satisfy
\begin{align}
\{\fb, \fbd\} = -1.
\end{align}
Then the states with odd number of $\fb$ will have negative (squared) norms. For such fermions the operator counting the
number of them in a state has to be defined as
\begin{align}
n_{\fb} = -\fbd \fb.
\end{align}

We also want the fermions to satisfy the same commutation relations (\ref{comrelbar}) with ${\bar T}$ as the bosons. This is achieved by the following transfer matrix:
\begin{align}
\label{Tbar}
{\bar T} & = \, :\exp \biggl(-{\beta \over \al} (\fbdown \fbdup + \bbdown \bbdup) + {\be^* \over \al^*}
(\fbup \fbddown + \bbup \bbddown) \biggr)\!: \nonumber \\
& \quad \times (\al^*)^{n_{\fb\up}+n_{\bb\up}} \al^{n_{\fb\down}+n_{\bb\down}}.
\end{align}
Notice that the bosonic part of this operator is the same as Eq. (\ref{Tbbar}). It is now easy to check that the commutators with bosons and fermions have the same form:
\begin{align}
{\bar T} \cbd_\sigma &= \cbd_{\sigma'} {\cal S}^\dagger_{\sigma'\sigma} {\bar T}, &
{\bar T} \cb_\sigma &= {\cal S}_{\sigma\sigma'} \cb_{\sigma'} {\bar T}.
\end{align}

As on the up links, these relations imply that the fermions and bosons on the down links transform as SU(2) spinors under commutation with ${\bar T}$. Their singlet bilinear combinations again form the generators of the $\mathfrak{osp}(2|2)$ superalgebra, and we define them as
\begin{align}
\label{down-gen}
\Bb &= -\frac{1}{2}(\bbdup \bbup + \bbddown \bbdown + 1),
& \Qb_3 &= \frac{1}{2}(\fbdup \fbup + \fbddown \fbdown + 1), \no \\
\Qb_+ &= \fbdown \fbup,  & \Qb_- &= \fbdup \fbddown, \no \\
\Vb_+ &= - \frac{1}{\sqrt 2} (\bbup \fbdown - \bbdown \fbup),
& \Wb_- &= (\Vb_+)^\dagger, \no \\
\Vb_- &= \frac{1}{\sqrt 2} (\fbdup \bbup + \fbddown \bbdown),
& \Wb_+ &= - (\Vb_-)^\dagger.
\end{align}
These operators satisfy the same commutation relations (\ref{comm-rels}) as the ones on the up links.

The quadratic Casimirs for the dual superspins $\bar J$ and $\bar K$ are defined in the same way as for $J$ and $K$, see Eqs. (\ref{casimir-J}) and (\ref{casimir-K}). Using the quadratic Casimirs we can introduce the invariant products of superspins:
\begin{align}
J\cdot \Jb &\equiv \Jb\cdot J = C_2(J + \Jb) - C_2(J) - C_2(\Jb) \no \\
&= 2Q_3 \bar{Q}_3 - 2B\bar{B}  -  V_- \bar{W}_+ + W_+ \bar{V}_- \no \\
&\quad + V_+ \bar{W}_- - W_- \bar{V}_+ + Q_+ \bar{Q}_- + Q_-\bar{Q}_+, \\
K \cdot \bar{K} &\equiv \bar{K}\cdot K = C_2(K + \bar{K}) - C_2(K) - C_2(\bar{K}) \no \\
&= 2Q_3 \bar{Q}_3 - 2B\bar{B}  -  V_- \bar{W}_+ + W_+ \bar{V}_-.
\end{align}

The transfer matrix ${\bar T}$ averaged over random SU(2) scattering matrices again gives the projector
\begin{align}
\la \bar{T} \ra_{\text{SU(2)}} = \bar{P}.
\label{proj-bar}
\end{align}
onto the space of SU(2) singlets $|{\bar m}\ra$, $m = 0, 1, 2$:
\begin{align}
|{\bar 0}\ra &= |\text{vacuum}\ra,  \\
|{\bar 1}\ra &= - {\bar W}_- |{\bar 0}\ra = \frac{1}{\sqrt 2}(\bbdup \fbddown - \bbddown \fbdup)|{\bar 0}\ra, \\
|{\bar 2}\ra &= - {\bar Q}_- |0\ra = - \fbdup \fbddown |{\bar 0}\ra.
\end{align}
These singlets form the representation $\pi(-\tfrac{1}{2},\tfrac{1}{2})$ of the $\mathfrak{osp}(2|2)$ algebra dual to the fundamental $\pi(\tfrac{1}{2},\tfrac{1}{2})$. Note that the state $|{\bar 1}\ra$ contains odd number of fermions, and,
therefore, has negative square norm:
\begin{align}
\la {\bar 1}|{\bar 1}\ra = -1.
\end{align}

The action of the generators on the states in the representation
$\pi(-\tfrac{1}{2},\tfrac{1}{2})$ is easily found to be
\begin{align}
{\bar Q}_3|{\bar 0}\ra &= \frac{1}{2}|{\bar 0}\ra, & {\bar Q}_3|{\bar 1}\ra &= 0, &
{\bar Q}_3|{\bar 2}\ra &= -\frac{1}{2}|{\bar 2}\ra, \nonumber \\
{\bar B}|{\bar 0}\ra & = -\frac{1}{2}|{\bar 0}\ra, & {\bar B}|{\bar 1}\ra &= -|{\bar 1}\ra, &
{\bar B}|{\bar 2}\ra &= -\frac{1}{2}|{\bar 2}\ra, \nonumber \\
{\bar Q}_+|{\bar 2}\ra & =  -|{\bar 0}\ra, & {\bar Q}_-|{\bar 0}\ra &= -|{\bar 2}\ra, \nonumber \\
{\bar V}_+|{\bar 1}\ra & = |{\bar 0}\ra, & {\bar V}_-|{\bar 1}\ra &= -|{\bar 2}\ra, \nonumber \\
{\bar W}_+|{\bar 2}\ra & = -|{\bar 1}\ra, & {\bar W}_-|{\bar 0}\ra &= -|{\bar 1}\ra.
\label{J-bar-action}
\end{align}

\subsection{Nodal transfer matrices}
\label{app:node-T}

With our choice of the scattering at the nodes to be diagonal in the spin index, the node evolution operators $T_{i,i+1}$ are simple generalizations of the ones used in the SUSY formulation of the CC model. Essentially, we just have to drop the advanced particles and take the product over the spin indices. As we mentioned in Sec. \ref{sec:SUSY}, the phase and the damping factors $e^{i\phi} = e^{i\delta- \eta}$ have been moved to the nodes, so we use the nodal scattering matrices (\ref{node-matrices-new}). This gives the following expression for $T_{12}$:
\begin{align}
T_{12} &=  \prod_{\sigma = \up,\down}
\Bigl[  \exp \Bigl( t_{A\sigma} e^{i\bar{\phi}_2} \bigl(\fd_{1\sigma} \fbd_{2\sigma} +
\bd_{1\sigma} \bbd_{2\sigma} \bigr) \Bigr) \no \\
&\quad \times \bigl(r_{A\sigma} e^{i\phi_1} \bigr)^{n_{f_{1\sigma}} +
n_{b_{1\sigma}}}  \bigl(r_{A\sigma} e^{i\bar{\phi}_2} \bigr)^{n_{\fb_{2\sigma}} + n_{\bb_{2 \sigma}}} \nonumber \\
&\quad \times \exp \Bigl(- t_{A\sigma}e^{i\phi_1} \bigl( \fb_{2\sigma} f_{1\sigma} +
\bb_{2\sigma} b_{1\sigma} \bigr) \Bigr) \Bigr],
\label{VSQH}
\end{align}
and a similar expression for $T_{23}$ (obtained by replacing all subscripts 1 by 3, and the changing the sublattice index $A$ to $B$).

We now simplify notation by dropping the site indices, since the fermions and bosons on the two sites (as well as the phases $\phi$) are differentiated by the overbar. Likewise, we drop the sublattice index. Then we can rewrite the evolution operators for both sublattices as:
\begin{align}
\label{T-product}
T &= T_+ T_0 {\bar T}_0 T_-, \\
T_+ &=  \prod_\sigma e^{t_{\sigma} e^{i\bar{\phi}} A^\dagger_\sigma}, & A^\dagger_\sigma &= \bd_\sigma \bbd_\sigma + \fd_\sigma \fbd_\sigma, \\
T_0 &= \prod_\sigma \bigl(r_{\sigma} e^{i\phi} \bigr)^{n_{f_\sigma} + n_{b_\sigma}}, &
{\bar T}_0 &= \prod_\sigma \bigl(r_{\sigma} e^{i\bar{\phi}} \bigr)^{n_{\fb_\sigma} + n_{\bb_\sigma}}, \\
T_- &= \prod_\sigma e^{- t_{\sigma} e^{i\phi} A_\sigma}, & A_\sigma &= \bb_\sigma b_\sigma + \fb_\sigma f_\sigma.
\end{align}

We need to project $T$ to the the tensor product $\pi(\tfrac{1}{2},\tfrac{1}{2}) \otimes \pi(-\tfrac{1}{2}, \tfrac{1}{2})$. Let us now denote the projection operator by ${\hat P} \equiv P \otimes {\bar P}$. We note, first of all, that due to irreducibility of $\pi(\pm\tfrac{1}{2},\tfrac{1}{2})$, the projected transfer matrix must be a linear combination of products of superspin components $J$ or $\Jb$ (and identity operators) on the two sites. Next we note that when projecting $T_\pm$, we need to expand the exponentials in $T_\pm$ only to linear order in each $A$ and $A^\dagger$ (since higher orders only contain triplet combinations of bosons and fermions on each site, and would take us out of the spaces of interest):
\begin{widetext}
\begin{align}
{\hat P} T {\hat P}
&= {\hat P} \big(1 + t_\up e^{i\bar{\phi}} A^\dagger_\up + t_\down e^{i\bar{\phi}} A^\dagger_\down + t_\up t_\down e^{2 i\bar{\phi}} A^\dagger_\up A^\dagger_\down\big)
T_0 {\bar T}_0 \big(1 - t_\up e^{i \phi} \Aph_\up - t_\down e^{i \phi} \Aph_\down + t_\up t_\down e^{2i\phi} \Aph_\up \Aph_\down \big)  {\hat P}.
\end{align}
$T_0$ and ${\bar T}_0$ act diagonally in the respective irreps, and can be written as $1 + \big(r_\up r_\down e^{2i\phi} - 1\big) (B + Q_3)$ and $1 - \big(r_\up r_\down e^{2i{\bar \phi}} - 1\big) ({\bar B} + {\bar Q}_3)$, respectively. Then when we develop the products in the last expression, linear and cubic terms in $A$ and $A^\dagger$ do not contribute, and neither do products of the type $A^\dagger_\sigma T_0 {\bar T}_0 \Aph_{-\sigma}$, so we have
\begin{align}
{\hat P} T {\hat P}  &= {\hat P} \big(T_0 {\bar T}_0 - t_\up^2 e^{i(\bar{\phi} + \phi)} A^\dagger_\up T_0 {\bar T}_0 \Aph_\up
- t_\down^2 e^{i(\bar{\phi} + \phi)} A^\dagger_\down T_0 {\bar T}_0 \Aph_\down \nonumber \\
& \quad + t_\up t_\down e^{2i\bar{\phi}} A^\dagger_\up A^\dagger_\down T_0 {\bar T}_0 + t_\up t_\down e^{2 i \phi} T_0 {\bar T}_0 \Aph_\up \Aph_\down
+ t_\up^2 t_\down^2 e^{2i(\bar{\phi} + \phi)} A^\dagger_\up A^\dagger_\down T_0 {\bar T}_0 \Aph_\up \Aph_\down \big){\hat P}.
\label{PTP-1}
\end{align}
Let us number terms in this expression (1) through (6), and consider them one by one. First we have
\begin{align}
\label{PT_0P}
(1)
&= 1 + \big(r_\up r_\down e^{2i\phi} - 1\big) (B + Q_3) - \big(r_\up r_\down e^{2i{\bar \phi}} - 1\big)({\bar B} + {\bar Q}_3)
- \big(r_\up r_\down e^{2i\phi} - 1\big)\big(r_\up r_\down e^{2i{\bar \phi}} - 1\big) (B + Q_3)({\bar B} + {\bar Q}_3).
\end{align}
We rearrange the next two terms in Eq. (\ref{PTP-1}) noticing that commutation of $T_0$ or ${\bar T}_0$ with fermions or bosons changes one of the number operators by one. Then we have
\begin{align}
{\hat P} A^\dagger_\sigma T_0 {\bar T}_0 \Aph_\sigma {\hat P}
&= \frac{e^{-i(\phi + {\bar \phi})}}{r_\sigma^2} {\hat P} \big(\bd_\sigma \bph_\sigma T_0 {\bar T}_0 \bbd_\sigma \bbph_\sigma
- \bd_\sigma \fph_\sigma T_0 {\bar T}_0 \bbd_\sigma \fbph_\sigma
+ \fd_\sigma \bph_\sigma T_0 {\bar T}_0 \fbd_\sigma \bbph_\sigma
+ \fd_\sigma \fph_\sigma T_0 {\bar T}_0 \fbd_\sigma \fbph_\sigma \big) {\hat P}.
\end{align}
Here we need to represent bilinears in bosons and fermions on each site as linear combinations of a singlet and a triplet bilinear, and then the projection operators $\hat P$ allow us to drop the triplets. This gives
\begin{align}
{\hat P} A^\dagger_\sigma T_0 {\bar T}_0 \Aph_\sigma {\hat P} &= \frac{e^{-i(\phi + {\bar \phi})}}{r_\sigma^2} {\hat P}
\Big[\Big(Q_3 + \frac{1}{2} \Big) T_0 {\bar T}_0 \Big(\Qb_3 - \frac{1}{2} \Big) -\Big(B - \frac{1}{2}\Big)T_0 {\bar T}_0 \Big(\Bb + \frac{1}{2} \Big)
- \frac{1}{2} V_- T_0 {\bar T}_0 \Wb_+ + \frac{1}{2} W_+ T_0 {\bar T}_0 \Vb_- \Big] {\hat P}.
\end{align}
This expression contains products of superspin components on each site. Due to irreducibility of $\pi(\pm\tfrac{1}{2},\tfrac{1}{2})$, such products can be replaced by linear combinations of superspin components. The easiest way to find these combinations is to use the matrix representations of the superspin components. The result is
\begin{align}
{\hat P} A^\dagger_\sigma T_0 {\bar T}_0 \Aph_\sigma {\hat P} &= e^{i(\phi + {\bar \phi})} r_{-\sigma}^2
\Big[\Big(Q_3 + \frac{1}{2} \Big) \Big(\Qb_3 - \frac{1}{2} \Big) - \Big(B - \frac{1}{2}\Big) \Big(\Bb + \frac{1}{2} \Big) -\frac{1}{2} V_- \Wb_+ + \frac{1}{2} W_+ \Vb_- \Big],
\end{align}
and
\begin{align}
(2) + (3) &= (2t_\up^2  t_\down^2 - t_\up^2 - t_\down^2) e^{2i(\phi + {\bar \phi})}
\Big[\Big(Q_3 + \frac{1}{2} \Big) \Big(\Qb_3 - \frac{1}{2} \Big) - \Big(B - \frac{1}{2}\Big) \Big(\Bb + \frac{1}{2} \Big)
- \frac{1}{2} V_- \Wb_+ + \frac{1}{2} W_+ \Vb_- \Big].
\label{2+3}
\end{align}

In the same way we treat the other terms:
\begin{align}
{\hat P} A^\dagger_\up A^\dagger_\down  T_0 {\bar T}_0  {\hat P}
&=
{\hat P} \big( -\fdup \fddown \fbdup \fbddown + \fdup \bddown \fbdup \bbddown + \bdup \fddown \bbdup  \fbddown
+ \bdup \bddown \bbdup \bbddown \big)  T_0 {\bar T}_0  {\hat P}
= - Q_+ \Qb_- - V_+ \Wb_-, \nonumber \\
{\hat P} T_0 {\bar T}_0 \Aph_\up \Aph_\down {\hat P} &= - Q_- \Qb_+ + W_- \Vb_+,
\end{align}
and
\begin{align}
(4) + (5) &= -t_\up t_\down e^{2i\bar{\phi}} (Q_+ \Qb_- + V_+ \Wb_-) - t_\up t_\down e^{2 i \phi} (Q_- \Qb_+ - W_- \Vb_+).
\end{align}

Finally, we need
\begin{align}
{\hat P} A^\dagger_\up A^\dagger_\down T_0 {\bar T}_0 \Aph_\up \Aph_\down {\hat P}
&= Q_+ Q_- \Qb_- \Qb_+ - Q_+ W_- \Qb_- \Vb_+  + V_+ Q_- \Wb_- \Qb_+ + V_+ W_- \Wb_- \Vb_+ \nonumber \\
&= 3 \Big(B - \frac{1}{2}\Big)\Big(\Bb + \frac{1}{2}\Big) - \Big(Q_3 + \frac{1}{2}\Big)\Big(\Qb_3 - \frac{1}{2}\Big)
- W_+ \Vb_- + V_- \Wb_+
\nonumber \\ & \quad
+ \Big(Q_3 + \frac{1}{2}\Big)\Big(\Bb + \frac{1}{2}\Big) + \Big(B - \frac{1}{2}\Big)\Big(\Qb_3 - \frac{1}{2}\Big).
\end{align}
The first four terms here are very similar to Eq. (\ref{2+3}), which leads to
\begin{align}
(2)+(3)+(6)
&=  e^{2i(\bar{\phi} + \phi)} \Big[t_\up^2 t_\down^2 (Q_3 + B)(\Qb_3 + \Bb)
- \frac{t_\up^2 + t_\down^2}{2} (K \cdot {\bar K} - B - Q_3 + \Bb + \Qb_3) \Big].
\end{align}

Collecting all six terms together, we arrive at Eq. (\ref{withRandomPhase}).
\end{widetext}

\bibliography{mybib}
\end{document}